\documentclass[12pt]{article}%
\usepackage{amssymb,amsbsy}
\usepackage{url}%

\newcommand{\be}{\begin{equation}}
\newcommand{\ee}{\end{equation}}
\newcommand{\ba}{\begin{array}}
\newcommand{\ea}{\end{array}}
\newcommand{\p}{\partial}
\newcommand{\fl}{}
\newcommand{\ds}{\displaystyle}

\begin{document}
\date{July 12, 2008}
\title%
{\protect\vspace*{-21mm} \bf A Coordinate-Free Construction
for a Class of Integrable Hydrodynamic-Type Systems}
\author{Maciej B\l aszak$^a$ and Artur Sergyeyev$^b$\\ 
$^a$Department of Physics, A. Mickiewicz University\\ Umultowska
85, 61-614 Pozna\'{n}, Poland\\ $^b$Mathematical Institute,
Silesian University in Opava,\\ Na Rybn\'\i {}\v{c}ku 1, 746\,01
Opava, Czech Republic\\ E-mail: {\tt blaszakm@amu.edu.pl} and {\tt
Artur.Sergyeyev@math.slu.cz}}
\maketitle

\begin{abstract}\vspace{-15mm}
Using a (1,1)-tensor $L$
with zero Nijenhuis torsion and maximal possible number (equal to
the number of dependent variables)
of distinct, functionally independent eigenvalues we define, in a coordinate-free fashion,
the seed systems which are weakly nonlinear semi-Hamiltonian systems
of a special form, and an infinite set of conservation laws for the seed systems.

The reciprocal transformations constructed from these conservation
laws yield a considerably larger class of hydrodynamic-type
systems from the seed systems, and we show that these new systems
are again defined in a coordinate-free manner, using the tensor
$L$ alone, and, moreover, are weakly nonlinear and
semi-Hamiltonian, so their general solution can be obtained by
means of the generalized hodograph method of Tsarev.

\looseness=-2
\end{abstract}

\section*{Introduction}

In the present paper we deal with the systems of first order
quasi-linear PDEs of the form \be\label{hds00}
\boldsymbol{u}_t=A(\boldsymbol{u})\boldsymbol{u}_x,\quad \quad \ee
where $\boldsymbol{u}=(u^1,\dots,u^n)^T$, $A$ is an $n\times n$
matrix, the superscript $T$ indicates the transposed matrix. The
systems (\ref{hds00}) are usually called \emph{hydrodynamic-type}
systems or \emph{dispersionless} systems. More specifically, we
shall restrict ourselves to considering the systems (\ref{hds00})
which are semi-Hamil\-tonian in the sense of Tsarev \cite{tsarev}
{\em and} weakly nonlinear\footnote{Note that weakly nonlinear
systems are also known as {\em linearly degenerate}, see e.g.\
\cite{serre,fdup,fp2,m}.} \cite{f1}.

Although the class of weakly nonlinear semi-Hamiltonian (WNSH)
systems was extensively studied in the literature, see e.g.\
\cite{m, f1,f2,fp2,gr,mvp} and references therein, the results
obtained so far were mostly presented in the distinguished
coordinates, the so-called Riemann invariants. In particular, in
these coordinates we have a complete description of WNSH
hydrodynamic-type systems \cite{f1,m} and, moreover, the general
solution in implicit form for any such system can be found. What
is more, any WNSH system written in the Riemann invariants can be
linearized using a suitably chosen reciprocal transformation
\cite{f1} (see e.g.\ \cite{kr1,kr2,cr1,fp2} and references therein
for a general theory of reciprocal transformations). \looseness=-1

However, not much is known so far
about how to construct or identify WNSH systems
in a coordinate-free fashion or construct reciprocal transformations
for such systems written in arbitrary coordinates.
Even though there exists \cite{pss} a coordinate-free version of
conditions under which a given hydrodyn\-am\-ic-type system is
weakly nonlinear and semi-Hamiltonian, the conditions in question
written in the coordinate-free form are quite cumbersome, and
constructing any reasonably large classes of WNSH systems in
arbitrary coordinates using these conditions is a virtually
impossible task even for low values of $n$ except for the simplest
cases of $n=2$ and $n=3$. As for the case of arbitrary $n$, some
results were obtained in \cite{m} for a special class of the WNSH
systems, namely, the seed systems, see below for details. \looseness=-1

In the present paper we construct in a coordinate-free fashion
fairly extensive classes of WNSH systems from the so-called seed systems.
We start with a (1,1)-tensor $L$
with zero Nijenhuis torsion and maximal possible number
(equal to the number of dependent variables)
of distinct, functionally independent eigenvalues. Using this tensor we
define, in a coordinate-free fashion,
a class of WNSH hydrodynamic-type systems which we call the {\em seed
systems}, see Section~\ref{dkslt} below for details.

We then observe that the seed systems possess infinitely many
nontrivial conservation laws of a special form that {\em can be
written in a coordinate-free fashion}. Note that even
though any semi-Hamiltonian system has infinitely many conservation
laws \cite{tsarev}, in general there is no way to write them down
explicitly in {\em arbitrary} coordinates. \looseness=-1

Using the above special conservation laws we construct the
reciprocal transformations (\ref{rct}) for the seed systems
and show that these transformations yield new large classes (\ref{kitr}) of
WNSH hydrodynamic-type systems {\em a priori} written in a
coordinate-free fashion. Finally, using the explicit form of the resulting systems in the
Riemann invariants, we write down general solutions for the systems
in question using the technique from \cite{f1,m}, see
Section~\ref{secgs} below for details.\looseness=-1

It is important to stress that, as shown
in Section~\ref{dksltrt} below,
for writing down the reciprocal transformations in question
it suffices to know the tensor $L$ alone. Thus,
the coordinate-free construction of weakly nonlinear
semi-Hamiltonian hydrodynamic-type systems laid out in the present paper
works for any (1,1)-tensor with zero Nijenhuis torsion
and maximal possible number of distinct, functionally independent eigenvalues.
Moreover, as shown in Section~\ref{dkslt} below,
this tensor always admits an infinite family of metrics
for which it is an $L$-tensor in the sense of \cite{ben93, ben97,ben05}.




\section{The seed systems}\label{dkslt}

Consider an $n$-dimensional manifold $M$ endowed with a tensor\footnote{For the sake of brevity
in what follows we shall use the term `tensor' instead of `tensor field'.}
$L$ of type (1,1), i.e., with one covariant and one contravariant index,
with zero Nijenhuis torsion and $n$ distinct, functionally independent
eigenvalues.  It can be shown that a tensor with these properties always is
an $L$-tensor \cite{ben93, ben97,ben05}, also known as a
special conformal Killing tensor of trace type \cite{c}.


Following \cite{ben93, ben05, mac2005}, consider the following set of tensors
of type (1,1) on $M$: \looseness=-1
\begin{equation}
K_{1}=\mathbb{I},\quad
K_{r}=\sum_{k=0}^{r-1}\rho_k L^{r-1-k},\quad
r=2,\dots,n,\label{Krec}%
\end{equation}
where 
$\mathbb{I}$ is the $n\times n$ unit matrix, and
$\rho_i$ are coefficients of the characteristic polynomial of the
tensor $L$, i.e.,
\begin{equation}
\det(\xi\mathbb{I}-L)=\sum\limits_{i=0}^{n}\rho_i \xi
^{n-i}.\label{Viete}
\end{equation}

Now consider a vicinity $U\subset M$ with local coordinates
$u^1,\dots,u^n$, and
a set of hydrodynamic-type systems of the form
\begin{equation}
\label{ki0}K_1^{-1} \boldsymbol{u}_{t_{1}}=K_2^{-1} \boldsymbol{u}_{t_{2}}
=\cdots=K_n^{-1} \boldsymbol{u}_{t_{n}}
\end{equation}
where let $\boldsymbol{u}=(u^1,\dots,u^n)^T$,
and the superscript $T$ refers to the matrix transposition,
$t_i$ are independent variables,
$K_i^{-1}$ are tensors of type (1,1) such that
$K_i K_i^{-1}=\mathbb{I}$, $i=1,\dots,n$.

For any fixed $j\in\{1,\dots,n\}$ we can rewrite (\ref{ki0}) as
\be \label{ki}
\boldsymbol{u}_{t_{i}}=K_{i} K_j^{-1}
\boldsymbol{u}_{t_j},\quad i=1,\dots,n,\quad i\neq j. \ee
Notice
that in (\ref{ki}) the variable $t_j$ plays the role of a space
variable while the remaining times $t_i$ should be considered as
evolution parameters. Moreover
$K_{i} K_j^{-1}$ again is a tensor of
type $(1,1)$. It is important to stress that the set (\ref{ki0}) (or (\ref{ki}))
of hydrodynamic-type systems is covariant under arbitrary changes of local coordinates on $M$,
and in fact the systems in question are well-defined on the whole of $M$.

We shall refer to the systems (\ref{ki0}) or (\ref{ki}) with $K_i$ given by (\ref{Krec})
as to the {\em seed systems}. In fact, these systems belong to a broader class
of the so-called dispersionless Killing systems \cite{blasak}.
It can be shown \cite{m} that the seed systems are weakly nonlinear and semi-Hamiltonian.

It is immediate from (\ref{Krec}) that if we choose the eigenvalues
$\lambda^i$, $i=1,\dots,n$, of $L$ for the local
coordinates on $U$ (this is possible because the Nijenhuis torsion
of $L$ vanishes and the eigenvalues in question are simple and
functionally independent), the quantities (\ref{Krec}) will be
diagonal in these coordinates, and thus the eigenvalues in question
will provide the Riemann invariants for the seed 
systems (\ref{ki}).
As will be shown in
Section 5,
the solution for these systems, when expressed using the Riemann invariants,
has the form
(\ref{gstrb}). Note that the Lax representations for these systems
also appear in the context of the so-called universal hierarchy
\cite{as1,as2}.


Interestingly enough, for the seed systems
we have \cite{blasak} an infinite
set of conservation laws that can be constructed in a {\em
coordinate-free fashion}.

In order to write this set down we 
need the so-called {\em basic separable potentials} $V_{r}^{(k)}$
that can be defined using the tensor $L$ via the following
recursion relation~\cite{mac2005}: \looseness=-2
\begin{equation}
V_{r}^{(k)}=V_{r+1}^{(k-1)}-\rho_r V_{1}^{(k-1)},
k\in\mathbb{Z},\label{rekup}
\end{equation}
with the initial condition%
\begin{equation}
V_{r}^{(0)}=-\delta_{r}^n,\qquad r=1,\ldots,n.\label{in}%
\end{equation}
Here and below we tacitly assume that $V_{r}^{(k)}\equiv 0$ for $r<1$ or $r>n$.

The recursion (\ref{rekup}) can be reversed. The inverse recursion is given by%
\begin{equation}
\fl V_{r}^{(k)}=V_{r-1}^{(k+1)}-\frac{\rho_{r-1}}{\rho_{n}} V_{n}^{(k+1)},\qquad
k\in
\mathbb{Z},\quad r=1,\ldots,n. \label{rekdown}%
\end{equation}
Hence, the first nonconstant potentials are $V_{r}^{(n)}=\rho_r$ for $k>0$
and $V_{r}^{(-1)}=\frac{\rho_{r-1}}{\rho_{n}}$ for $k<0$, respectively.

The conservation laws in question read \cite{blasak}
\be\label{cl}
D_{t_i} (V_j^{(k)})=D_{t_j}(V_i^{(k)}),\quad i,j=1,\dots,n,\quad i\neq j,\quad k\in\mathbb{Z},
\ee
where $D_{t_i}$ are total derivatives computed by virtue of (\ref{ki0}).
These conservation laws are obviously nontrivial for all integer $k\neq 0,\dots,n-1$.

\section{Reciprocal transformations and more}\label{dksltrt}
Using (\ref{cl}) we can define a large class of reciprocal
transformations for the seed systems. Using these transformations we
construct extensive new classes (\ref{kitr})
of WNSH hydrodynamic-type systems. Most importantly, these
transformed systems, just like their seed counterparts,
possess an infinite set of nontrivial conservation
laws that can be constructed in a coordinate-free fashion, and the
general solution of any of the transformed systems (\ref{kitr}) written in the
Riemann invariants takes the form (\ref{gst1}) and (\ref{gst2}).
\looseness=-1

The reciprocal transformation in question is defined for the whole
set (\ref{ki0}) of the seed systems and reads \cite{am07} as follows:
\begin{equation}
\begin{array}{l}
\label{rct}\displaystyle d\tilde t_{s_i}=-\sum\limits_{j=1}^n
V_j^{(\gamma_i)} dt_{j},\quad i=1,\dots,k,\\
\tilde t_{m}=t_{m},\quad m=1,2,\dots,n,\quad m\neq s_a\quad\mbox{for
any}\quad a=1,\dots,k.
\end{array}
\end{equation}

Here $1\leq k\leq n$; the numbers $s_a$, $a=1,\dots,k$, are a $k$-tuple of
distinct integers from the set $\{1,\dots,n\}$,
and $\gamma_j$ are arbitrary positive integers that satisfy
the following conditions:
\be\label{gamma}
\gamma_1>\gamma_2>\cdots>\gamma_k>n-1.
\ee
The choice of numbers $k\in\{1,\dots,n\}$, $s_a$, and $\gamma_a$
that satisfy the above conditions uniquely determines the transformation (\ref{rct}).
Using (\ref{cl}) we can readily check that (\ref{rct}) is a well-defined reciprocal transformation.

The inverse of (\ref{rct}) has the form
\begin{equation}
\begin{array}{l}
\label{rcti}\displaystyle d t_{s_i}=-\sum\limits_{j=1}^n
\tilde V_j^{(n-s_i)} d\tilde t_{j},\quad i=1,\dots,k,\\
t_{l}=\tilde t_{l},\quad q=1,2,\dots,n,\quad l\neq s_a\quad\mbox{for
any}\quad a=1,\dots,k.
\end{array}
\end{equation}
Here $\tilde V_{j}^{(m)}$ are
{\em deformed separable potentials}
defined for all integer $m$
as follows:\\
1) for $j=s_1,\dots,s_k$ we define $\tilde V_{s_i}^{(m)}$ by means of the relations
\begin{equation}\label{v1}
V_{s_i}^{(m)}+\sum_{p=1}^k \tilde
V_{s_p}^{(m)}V_{s_i}^{(\gamma_p)}=0,
\end{equation}
whence
\begin{equation}\label{dual1linv}
\tilde V_{s_i}^{(m)}=-\det W_i^{(m)}/\det W,
\end{equation}
where $W$ is a $k\times k$ matrix of the form
\be\label{W}
W=\left|\!\left|\begin{array}{ccc}
V_{s_1}^{(\gamma_1)} & \cdots & V_{s_1}^{(\gamma_k)}\\
\vdots & \ddots & \vdots\\
V_{s_k}^{(\gamma_1)} & \cdots & V_{s_k}^{(\gamma_k)}
\end{array}\right|\!\right|,
\ee
and $W_i^{(m)}$ are obtained from $W$ by replacing
$V_{s_j}^{(\gamma_i)}$ by $V_{s_j}^{(m)}$ for all
$j=1,\dots,k$;\\
2) for $j\neq s_1,\dots,s_k$ we set
\be\label{v2} \tilde
V_j^{(m)}=V_j^{(m)}+\sum_{p=1}^k \tilde
V_{s_p}^{(m)}V_j^{(\gamma_p)},
\ee
or equivalently
\begin{equation}\label{dual1linv2}
\tilde V_{j}^{(m)}=\det \hat W_j^{(m)}/\det W,
\end{equation}
where $\hat W_j^{(m)}$ is a $(k+1)\times (k+1)$ matrix of the form
\be\label{hatW}
\hat W_j^{(m)}=\left|\!\left|\begin{array}{cccc}
V_{j}^{(m)} & V_{j}^{(\gamma_1)} & \cdots & V_{j}^{(\gamma_k)}\\
V_{s_1}^{(m)} & V_{s_1}^{(\gamma_1)} & \cdots & V_{s_1}^{(\gamma_k)}\\
\vdots & \vdots & \ddots & \vdots\\
V_{s_1}^{(m)} & V_{s_k}^{(\gamma_1)} & \cdots & V_{s_k}^{(\gamma_k)}
\end{array}\right|\!\right|.
\ee
It can be shown that the above definition of $\tilde V_i^{(j)}$ is equivalent
to the one given in \cite{am07}.

In order to find out how Eq.(\ref{ki0})
transforms under (\ref{rct}), we temporarily rewrite the former as
\be\label{kiaux}
\boldsymbol{u}_{t_{i}}=K_{i} \boldsymbol{Y},\quad i=1,\dots,n,
\ee
where $\boldsymbol{Y}$ is an arbitrary vector field on $M$.

The transformation (\ref{rct}) sends the set (\ref{kiaux}) of
seed systems into the following set:
\begin{equation}
\label{kiauxtr}
\boldsymbol{u}_{\tilde t_{i}}=\tilde K_{i}  \boldsymbol{Y},
\quad i=1,2,\dots,n,
\end{equation}
which upon elimination of $\boldsymbol{Y}$ can be written in the form similar to (\ref{ki0}):
\begin{equation}
\label{kitr}
\tilde K_1^{-1} \boldsymbol{u}_ {{\tilde t}_{1}}=\tilde K_2^{-1} \boldsymbol{u}_{{\tilde t}_{2}}
=\cdots=\tilde K_n^{-1} \boldsymbol{u}_{{\tilde t}_{n}},
\end{equation}
and can be further rewritten like (\ref{ki})
\be\label{kitr1}
\boldsymbol{u}_{\tilde t_{i}}=\tilde K_{i} \tilde K_j^{-1} \boldsymbol{u}_{\tilde t_j},
\quad i=1,\dots,n,\quad i\neq j,
\ee
for any fixed $j\in\{1,\dots,n\}$. As will be shown in Section 5, the general solution for these
systems is given by (\ref{gst1}) and (\ref{gst2}).

Using (\ref{rcti}) and the chain rule we find, after a straightforward but tedious computation,
that
\begin{equation}\label{ktr_a}\begin{array}{lll}
\displaystyle \tilde{K}_{s_i}&=&-\sum\limits_{j=1}^k
\tilde V_{s_i}^{(n-s_j)} K_{s_j}M^{-1},\quad i=1,\dots,k,\\[5mm]
\tilde{K}_{m}&=& K_{m}M^{-1}
-\sum\limits_{l=1}^k \tilde V_m^{(n-s_l)} K_{s_l}M^{-1},
\quad m=1,2,\dots,n,\quad m\neq s_a\quad\mbox{for
any}\quad a=1,\dots,k,
\end{array}
\end{equation}
where 
\be\label{btgj} M=-\det W_{s_1}/\det W,
\ee
$W$ is given by (\ref{W}), and $W_{s_1}$ is obtained from $W$
by replacing $V_{s_j}^{(\gamma_1)}$ by $K_{s_j}$ for all
$j=1,\dots,k$. Here $\det W_{s_1}$ is a formal determinant with
matrix-valued entries of the same kind as in \cite{mac2005}.

Likewise, from (\ref{rct}) we infer that
\begin{equation}\label{ktri}
\begin{array}{lll}
\displaystyle K_{s_i}&=&-\sum\limits_{j=1}^k
V_{s_i}^{(\gamma_j)} \tilde K_{s_j} \tilde M^{-1},\quad i=1,\dots,k,\\[5mm]
K_{m}&=&\tilde K_{m} \tilde M^{-1}-\sum\limits_{l=1}^k V_m^{(\gamma_l)}\tilde K_{s_l} \tilde M^{-1},
\quad m=1,2,\dots,n,\quad m\neq s_a\quad\mbox{for
any}\quad a=1,\dots,k.\end{array}
\end{equation}
Here 
\be\label{btgj2} \tilde
M=-\det\widetilde{W}_{s_1}/\det\widetilde{W}, \ee $\widetilde{W}$
is a $k\times k$ matrix of the form \be\label{tW}
\widetilde{W}=\left|\!\left|\begin{array}{ccc} \tilde
V_{s_1}^{(\gamma_1)} & \cdots & \tilde V_{s_1}^{(\gamma_k)}\\
\vdots & \ddots & \vdots\\ \tilde V_{s_k}^{(\gamma_1)} & \cdots &
\tilde V_{s_k}^{(\gamma_k)}
\end{array}\right|\!\right|,
\ee
and $\widetilde{W}_{s_1}$ is obtained from $\widetilde{W}$
by replacing $\tilde V_{s_j}^{(\gamma_1)}$ by $\tilde K_{s_j}$ for all
$j=1,\dots,k$.

Eq.(\ref{kitr}) possesses the following infinite set
of nontrivial conservation laws similar to (\ref{cl}):
\be\label{cltr}\vspace*{-2mm}
\ba{l}
D_{\tilde t_i} (\tilde V_j^{(m)})=D_{\tilde t_j}(\tilde V_i^{(m)}),
\quad i,j=1,\dots,n,\quad i\neq j,\\
\qquad m\in\mathbb{Z},\quad m\neq\gamma_l,\quad l=1,\dots,k,
\quad m\not\in(\{1,\dots,n\}/\{s_1,\dots,s_k\}),
\ea
\ee
where 
the derivatives $D_{\tilde t_i}$ are computed by virtue of (\ref{kitr}).

As we have already mentioned above, 
any tensor $L$ of type (1,1)
with zero Nijenhuis torsion and $n$ distinct, functionally independent
eigenvalues always is
an $L$-tensor for some family of metrics on $M$.
In fact, in the coordinate frame associated
with the eigenvalues $\lambda^i$, $i=1,\dots,n$, of $L$, the most general family
of such contravariant metrics is given by (\ref{metrsep}). The quantities $K_i$ (\ref{Krec})
are then Killing tensors of type (1,1) for any metric tensor from the family (\ref{metrsep}).

Using the results of \cite{mac2005} it can be shown that the quantities $\tilde K_i$
are Killing tensors of type (1,1) for a contravariant metric $MG$, where
$G$ is {\em any} contravariant metric from the family (\ref{metrsep}).
Thus Eq.(\ref{kitr}) (or equivalently Eq.(\ref{kitr1})) indeed defines a
set of dispersionless Killing systems, and the systems (\ref{kitr1})
are weakly nonlinear and semi-Hamiltonian. Note that the weak
nonlinearity of (\ref{kitr}) can also be inferred from the general
result of Ferapontov (Proposition 3.2 of \cite{fdup}) stating that
reciprocal transformations of hydrodynamic-type systems preserve
weak nonlinearity. Alternatively, one can readily verify
weak nonlinearity and semi-Hamiltonicity of
(\ref{kitr}) in the coordinate frame associated
with the eigenvalues $\lambda^i$, $i=1,\dots,n$, of $L$. Moreover, in the next section
we show how to construct a general solution for any system (\ref{kitr})
in this coordinate frame using the method from \cite{f1,m}.


\section{Weakly nonlinear semi-Hamiltonian systems
in Riemann invariants: general solution from separation
relations} \label{secgs} Consider a hydrodynamic-type system
written in the Riemann invariants: \be\label{hdsri}
\lambda^i_{t}=v^i(\boldsymbol{\lambda})\lambda^i_x,\quad
i=1,\dots,n, \ee where
$\boldsymbol{\lambda}=(\lambda^1,\dots,\lambda^n)$, and there is
no sum over $i$.

The system (\ref{hdsri}) is said to be {\em weakly nonlinear} (or
{\em linearly degenerate}, see e.g.\ \cite{f1,serre} and
references therein) if \be\label{wn} \p v^i/\p\lambda^i=0,\quad
i=1,\dots,n, \ee
and is said to be {\em semi-Hamiltonian} 
\cite{tsarev} if
\be\label{sh}
\frac{\p}{\p\lambda_j}\left(\frac{\p v^i/\p\lambda^k}{v^k-v^i}\right)
=\frac{\p}{\p\lambda_k}\left(\frac{\p v^i/\p\lambda^j}{v^j-v^i}\right),
\quad i,j,k=1,\dots,n,\quad i\neq j\neq k\neq i.
\ee

It is natural to ask which is the most general weakly nonlinear semi-Hamiltonian (WNSH)
hydrody\-na\-mic-type system (\ref{hdsri}) written in the Riemann invariants,
or, in other words, which is the most general form
of $v^i$ that satisfy (\ref{wn}) and (\ref{sh}).

It turns out \cite{f1,m} that any WNSH hydrodynamic-type system (\ref{hdsri})
admits $n-1$ commuting flows of the same kind, so we
actually have a {\em set} of commuting WNSH
hydrodynamic-type systems just like (\ref{ki}).\looseness=-1

In complete analogy with (\ref{ki}), this set can be written in a symmetric form as
\be\label{kiri0}
\frac{\lambda^i_{t_1}}{v_1^i}=\cdots
=\frac{\lambda^i_{t_n}}{v_n^i},\quad i=1,\dots,n,
\ee
where $v_1^i\equiv v^i$, $i=1,\dots,n$.
The most general form of such a set of WNSH hydrodynamic-type systems
is given by the formulas \cite{f1,m}
\begin{equation}\label{kif}
v_{r}^{i}=(-1)^{r+1}\frac{\det\Phi^{ir}}{\det\Phi^{i1}}.
\end{equation}
Here $\Phi$ is a matrix of the form \cite{f1,m}
\be\label{phi}
\Phi=\left(
\begin{array}{ccccc}
\Phi _{1}^{1}(\lambda^{1}) & \Phi _{1}^{2}(\lambda^{1}) & \cdots & \Phi
_{1}^{n-1}(\lambda^{1}) & \Phi_{1}^{n}(\lambda^{1}) \\
\vdots & \vdots & \cdots & \vdots & \vdots \\
\Phi _{n}^{1}(\lambda^{n}) & \Phi _{n}^{2}(\lambda^{n}) & \cdots & \Phi
_{n}^{n-1}(\lambda^{n}) & \Phi_{n}^{n}(\lambda^{n})%
\end{array}%
\right),
\end{equation}
where $\Phi _{j}^{i}(\lambda^{i})$ are arbitrary functions of the corresponding variables;
$\Phi^{ik}$ is the $(n-1)\times (n-1)$ matrix obtained from $\Phi$ by removing
its $i^{\mathrm{th}}$ row and $k^{\mathrm{th}}$ column. Note that we can,
without loss of generality, impose the normalization $\Phi_i^n=1$, $i=1,\dots,n$,
but we shall not use this normalization below.

The general solution for (\ref{kiri0})
can be written as \cite{f1,m}
\be\label{gstr}
\ds\sum\limits_{j=1}^n\ds\int^{\lambda^j}\ds\frac{\Phi_j^{n-r}(\xi)}{\varphi_j(\xi)} d\xi=t_r,\quad r=1,\dots,n,
\ee
where $\varphi_j(\xi)$ are arbitrary functions of a single variable.

If we fix $r,k\in\{1,\dots,n\}$, $r\neq k$, and consider the system
\be\label{wns}
{\lambda^i_{t_k}}=\frac{v_k^i}{v_r^i}{\lambda^i_{t_r}},\quad i=1,\dots,n,
\ee
then the general solution of (\ref{wns}) is given by (\ref{gstr}) with $t_j=\mathrm{const}$ for all $j\neq r,k$.
For any pair $(r,k)$ the system (\ref{wns}) represents (\ref{hdsri}), where $t_k=t$, $t_r=x$, and $v^i=v_k^i/v_r^i$
satisfy the conditions (\ref{wn}) and (\ref{sh}).
\looseness=-1

Note that to a given matrix $\Phi$ (\ref{phi}), or, equivalently, to
a set of $n$ Killing tensors and a class of metrics that admit
them, we can associate the so-called {\em separation relations} of
the form \be\label{sr}
\sum\limits_{j=1}^n\Phi_i^j(\lambda^i)H_j=f_i(\lambda^i)\mu_i^2,
\quad i=1,\dots,n, \ee where $H_j$ are separable geodesic
Hamiltonians and $f_i(\xi)$, $i=1,\dots,n$, are arbitrary
functions of a single variable \cite{Sklyanin, m,am07} which are
related to their counterparts in (\ref{gstr}) via the formula
\be
\varphi_i(\xi)=\left(f_i(\xi)\sum\limits_{j=1}^n\Phi_i^j(\xi)a_j\right)^{1/2},
\ee
where $a_j$ are arbitrary constants.

From the point of view of separation relations (\ref{sr}) the
matrix $\Phi$ (\ref{phi}) is nothing but the St\"ackel matrix
related to the Hamiltonians $H_i$. Moreover,
the commutativity of $H_i$ implies the
commutativity of the associated flows (\ref{kiri0}),
and the general solution (\ref{gstr}) for (\ref{kiri0}) in fact
can be obtained \cite{f1,m} from the general solution
for the simultaneous equations of motion for $H_i$
which, in turn, is found using the separation relations (\ref{sr}).\looseness=-1

Let us briefly recall the rationale behind
the separation relations (\ref{sr}). 
If we define the Hamiltonians
$H_i=H_i(\boldsymbol{\lambda},\boldsymbol{\mu})$, $i=1,\dots,n$,
where $\boldsymbol{\mu}=(\mu_1,\dots,\mu_n)^T$, as solutions of
the system (\ref{sr}) of linear algebraic equations then these
Hamiltonians have the form
\be\label{hi}
H_i=\boldsymbol{\mu}^T K_i G \boldsymbol{\mu}=\sum\limits_{r,s=1}^n
\mu_r (K_i G)^{rs} \mu_s,\quad i=1,\dots,n,
\ee
and are well-known to
Poisson commute with respect to the canonical Poisson bracket
$\{\lambda^i,\mu_j\}=\delta_i^j$. Quite naturally, $K_i$ are
Killing tensors of type $(1,1)$
for a contravariant metric $G$. However, it is important to stress
that in general these $K_i$ do not necessarily
have the form (\ref{Krec}).

We know from \cite{am07} that for the seed systems
the separation relations (\ref{sr})
read \be\label{srb} \sum\limits_{j=1}^n (\lambda^i)^{n-j}
H_j=f_i(\lambda^i)\mu_i^2,\quad i=1,\dots,n, \ee while for the
systems (\ref{kitr}) we have
\begin{equation}\label{srd}
\sum\limits_{j=1}^k (\lambda^i)^{\gamma_j} \tilde H_{s_j}
+\sum\limits_{p=1, p\neq s_1,\dots,s_k}^n (\lambda^i)^{n-p} \tilde
H_p=f_i(\lambda^i)\mu_i^2,\quad i=1,\dots,n.
\end{equation}
Using (\ref{srb}) and (\ref{srd}) we can readily read off the
functions $\Phi_i^j$ from (\ref{phi}) associated with (\ref{ki0})
for $K_i$ given by (\ref{Krec}), and with (\ref{kitr}), and
construct general solution for any given seed system from (\ref{ki}),
and for the transformed systems
(\ref{kitr1}), by the method of \cite{f1,m}, see below for details.

For the special case of (\ref{srb}) the Killing tensors $K_i$
in (\ref{hi}) are given by (\ref{Krec}), and in the $\lambda$-coordinates $L$ has the form
\be\label{lsep}
L=\mathop{\rm diag}(\lambda^1,\dots,\lambda^n).
\ee
On the other hand, from the separation relations (\ref{srd})
we find that
\be\label{hi1}
\tilde H_i=\boldsymbol{\mu}^T \tilde{K}_i\tilde{G}\boldsymbol{\mu}=\sum\limits_{r,s=1}^n
\mu_r (\tilde{K}_i\tilde{G})^{rs} \mu_s,\quad i=1,\dots,n,
\ee
where $\tilde K_i$ are given by (\ref{ktr_a}) and $\tilde{G}=MG$ with $M$ given by (\ref{btgj}).
Thus, the Hamiltonian $H_i$ (resp.\ $\tilde H_i$) is naturally associated with the
twice contravariant Killing tensor $K_i G$ (resp.\ $\tilde{K}_i\tilde{G}$), and vice versa.

Now, the Hamiltonian $H_1$ associated with $K_1 G=G$, i.e., with the
original contravariant metric $G$ itself,
is the coefficient at the highest power of
$\lambda^i$ on the left-hand side of (\ref{srb}). Likewise, in view
of (\ref{gamma}) the coefficient at the highest power of $\lambda^i$
on the left-hand side of (\ref{srd}) equals $\tilde H_{s_1}$. This
is the reason why it is natural to consider the contravariant
metric $\tilde G$ associated with $\tilde H_{s_1}$
from (\ref{srd}) as a natural counterpart of the original contravariant
metric $G$,
cf.\ \cite{mac2005}.
\looseness=-1

Note also that the set of Hamiltonians $H_i$, $i=1,\dots,n$, in (\ref{srb}) is related to
the set of $\tilde H_i$, $i=1,\dots,n$,
in (\ref{srd}) via the so-called multiparameter generalized St\"ackel transform of a
special form, see \cite{am07} for further details, and this very fact uniquely determines the shape
of the reciprocal transformation (\ref{rct}) and (\ref{rcti}) relating (\ref{ki}) and (\ref{kitr}).
\looseness=-1

Let us now apply the above results on general solutions to the
seed systems (\ref{ki0}) and their transformed
counterparts (\ref{kitr}) in the
coordinates $\lambda^i$ being the eigenvalues of $L$.
In the coordinates in question (\ref{lsep}) holds
by assumption.

Note that any metric $G$ that admits $L$ of the form (\ref{lsep}) as
an   $L$-tensor in the $\lambda$-coordinates can be written in the
form
\cite{mac2005}%
\be\label{metrsep} G=\mathrm{diag}\left(
\frac{f_1(\lambda_1)}{\Delta_{1}},\ldots,
\frac{f_n(\lambda_n)}{\Delta_{n}}\right), \ee where $\Delta_{i}=
{\textstyle\prod\limits_{j\neq i}} (\lambda^{i}-\lambda^{j})$.
The class (\ref{metrsep}) with arbitrary functions $f_i(\xi)$ is
precisely the class of the metrics that admit the set of Killing
tensors given by (\ref{Krec}) with $L$ of the form (\ref{lsep}).

The joint general solution for the set of systems (\ref{ki0})
written in the Riemann invariants, that is,
\[
\frac{\lambda^i_{t_1}}{G^{ii} \p\rho_1/\p\lambda^i}=\cdots
=\frac{\lambda^i_{t_n}}{G^{ii} \p\rho_n/\p\lambda^i},\quad
i=1,\dots,n,
\]
or equivalently,
\be\label{kiribwr}
\frac{\lambda^i_{t_1}}{\p\rho_1/\p\lambda^i}=\cdots
=\frac{\lambda^i_{t_n}}{\p\rho_n/\p\lambda^i},\quad i=1,\dots,n,
\ee
where we used the formula
$\sum_{j=0}^{r-1} (\lambda^i)^{r-1-j} \rho_j=\p\rho_r/\p\lambda^i$ (see e.g.\ \cite{blasak}),
reads
\be\label{gstrb}
\ds\sum\limits_{j=1}^n\ds\int^{\lambda^j}\ds\frac{\xi^{n-r}}{\varphi_j(\xi)} d\xi=t_r,\quad r=1,\dots,n.
\ee
Notice that $\rho_{i}$ are nothing but the Vi\`{e}te polynomials in the variables $\lambda$.

Likewise, using (\ref{srd}) we see that the general solution of
(\ref{kitr}) in implicit form reads
\begin{eqnarray}\label{gst1}
\sum\limits_{j=1}^n\int^{\lambda^j} \ds\frac{\xi^{\gamma_q}}{\varphi_j(\xi)}d\xi
&=&\tilde t_{s_q},\quad q=1,\dots,k,\\
\label{gst2}\sum\limits_{j=1}^n\int^{\lambda^j}  \ds\frac{\xi^{n-i}}
{\varphi_j(\xi)}d\xi&=&\tilde t_i,\quad i=1,\dots,n,\quad i\neq s_q,\quad q=1,\dots,k.
\end{eqnarray}

\section*{Acknowledgments}

This research was supported in part
by the Ministry of Education, Youth and Sports of the Czech Republic
(M\v{S}MT \v{C}R) under grant MSM 4781305904, by the Ministry of
Science and Higher Education (MNiSW) of the Republic of Poland under
the research grant No.~N~N202~4049~33, and by Silesian University in
Opava under grant IGS 9/2008.
M.B. appreciates the warm hospitality of the Mathematical Institute of Silesian
University in Opava, where the present work was
initiated. A.S. is pleased to thank to
the Department of Physics of the Adam Mickiewicz University in Pozna\'n
for the warm hospitality
extended to him at the penultimate stage of preparation of the present paper.


\vspace*{-5mm}
\begin{thebibliography}{99}
{\setlength{\parskip}{1mm} \small
\bibitem{as1}L. Mart\'\i{}nez Alonso, A.B. Shabat, Towards a theory of differential constraints
of a hydrodynamic hierarchy. J. Nonlinear Math. Phys. \textbf{10} (2003), no. 2, 229--242;
preprint nlin.SI/0310036 (arXiv.org)

\bibitem{as2}L. Mart\'\i{}nez Alonso, A.B. Shabat,
Hydrodynamic reductions and solutions of the universal hierarchy.
Theoret. and Math. Phys. \textbf{140} (2004), no. 2, 1073--1085;
preprint nlin.SI/0312043 (arXiv.org)

\bibitem{ben93}S. Benenti, Orthogonal separable dynamical systems, in:
Differential geometry and its applications (Opava,~1992),
Silesian University in Opava, Opava, 1993,
pp.163--184; available online at
\url{http://www.emis.de/proceedings/5ICDGA/}

\bibitem{ben97}S. Benenti, Intrinsic characterization of the variable
separation in the Hamilton-Jacobi equation, J. Math. Phys.
\textbf{38} (1997), 6578--6602.

\bibitem{ben05} S. Benenti, Special symmetric two-tensors, equivalent
dynamical systems, cofactor and bi-cofactor systems, Acta Appl. Math.
\textbf{87 }(2005), 33--91.


\bibitem{m}M. B\l aszak and W.-X. Ma, Separable Hamiltonian equations
on Riemann manifolds and related integrable hydrodynamic systems, J.
Geom. Phys. \textbf{47} (2003), 21--42; preprint nlin.SI/0209014
(arXiv.org)


\bibitem{mac2005}M. B\l aszak, Separable systems with quadratic in
momenta first integrals, J. Phys. A: Math. Gen. \textbf{38} (2005),
1667--1685; preprint nlin.SI/0312025  (arXiv.org)
\looseness=-1

\bibitem{blasak}M. B\l aszak and K. Marciniak,
From St\"{a}ckel systems to integrable hierarchies of PDE's: Benenti
class of separation relations, J. Math. Phys. \textbf{47} (2006), paper 032904;
preprint nlin.SI/0511062 (arXiv.org)



\bibitem{c}M. Crampin and W. Sarlet, A class of non-conservative
Lagrangian systems on Riemannian manifolds, J. Math. Phys.
\textbf{42} (2001), 4313--4326.


\bibitem{f1}E.V. Ferapontov, Integration of weakly nonlinear
hydrodynamic systems in Riemann invariants, Phys. Lett. A
\textbf{158} (1991), 112--118.

\bibitem{fdup}E.V. Ferapontov, Dupin hypersurfaces and
integrable Hamiltonian systems of hydrodynamic type,
which do not possess Riemann invariants, Differential Geom. Appl.
{\bf 5} (1995), no. 2, 121--152.

\bibitem{f2}E.V. Ferapontov and A.P. Fordy, Separable Hamiltonians
and integrable systems of hydrodynamic type, J. Geom. Phys.
\textbf{21} (1997), 169--182.


\bibitem{fp2} E.V. Ferapontov and M.V. Pavlov, Reciprocal transformations
of Hamiltonian operators of hydrodynamic type: nonlocal
Hamiltonian formalism for linearly degenerate systems, J. Math.
Phys. \textbf{44} (2003),
1150--1172; preprint nlin.SI/0212026 (arXiv.org) \looseness=-1

\bibitem{gr} A.M. Grundland, M.B. Sheftel, P. Winternitz,
Invariant solutions of hydrodynamic-type equations. J. Phys. A:
Math. Gen. {\bf 33} (2000), no. 46, 8193--8215.









\bibitem{kr1}J.G.  Kingston, C. Rogers,
Reciprocal B\"acklund transformations of conservation laws. Phys.
Lett. A {\bf 92} (1982), no. 6, 261--264.

\bibitem{kr2}J.G.  Kingston, C. Rogers,
B\"acklund transformations for systems of conservation laws. Quart.
Appl. Math. {\bf 41} (1983/84), no. 4, 423--431.







\bibitem{mvp}M.V. Pavlov, Hamiltonian formalism of weakly nonlinear systems in hydrodynamics,
Theor. Math. Phys. \textbf{73} (1987), no. 2, 1242--1246.

\bibitem{pss} M.V. Pavlov, S.I. Svinolupov, R.A. Sharipov,
An invariant criterion for hydrodynamic integrability.
Funct. Anal. Appl. {\bf 30} (1996), no. 1, 15--22; preprint solv-int/9407003 (arXiv.org)





\bibitem {cr1}C. Rogers and W.F. Shadwick, {\em B\"acklund Transformations and
Their Applications}, Mathematics in Science and Engineering Series,
New York, Academic Press, 1982.




\bibitem{am07} A. Sergyeyev, M. B\l aszak, Generalized St\"ackel Transform and Reciprocal
Transformations for Finite-Dimensional Integrable Systems, J.
Phys. A: Math. Theor., {\bf 41} (2008), paper 105205;
preprint arXiv:0706.1473 [nlin.SI] (arXiv.org)

\bibitem{serre}D. Serre, {\em Systems of conservation laws. 1. Hyperbolicity, entropies, shock
waves.}
Cambridge, Cambridge University Press, 1999.

\bibitem{Sklyanin}E.K. Sklyanin, Separation of variables---new trends,
Progr. Theoret. Phys. Suppl. \textbf{118} (1995), 35--60; preprint
solv-int/9504001 (arXiv.org)



\bibitem{tsarev}S.P. Tsarev, The geometry of Hamiltonian systems of hydrodynamic type.
The generalized hodograph method.
{\em Math. USSR-Izv.} {\bf 37} (1991),  no. 2, 397--419.

%
%
%
%
}
\end{thebibliography}
\end{document}